\documentclass{article}
\usepackage{spconf,amsmath,graphicx}

\usepackage{enumitem}
\setlist{nosep, leftmargin=14pt}

\usepackage{mwe} 
\usepackage{amssymb}
\usepackage{xcolor,colortbl}
\usepackage{booktabs}
\usepackage{hyperref}
\usepackage{multirow}

\usepackage{enumitem}

\def\Q{\textbf{Q}}
\def\K{\textbf{K}}
\def\V{\textbf{V}}

\def\R{\mathbb{R}}

\definecolor{mygray}{gray}{0.9}

\title{Stroke Lesion Segmentation using Multi-Stage Cross-Scale Attention}
\name{Liang Shang, William A. Sethares, Anusha Adluru, Andrew L. Alexander}
\namee{Vivek Prabhakaran, Veena A. Nair, Nagesh Adluru}
\address{University of Wisconsin-Madison, Madison, WI, United States}
\begin{document}

\maketitle

\begin{abstract}
Precise characterization of stroke lesions from MRI data has immense value in prognosticating clinical and cognitive outcomes following a stroke. Manual stroke lesion segmentation is time-consuming and requires the expertise of neurologists and neuroradiologists. Often, lesions are grossly characterized for their location and overall extent using bounding boxes without specific delineation of their boundaries. While such characterization provides some clinical value, to develop a precise mechanistic understanding of the impact of lesions on post-stroke vascular contributions to cognitive impairments and dementia (VCID), the stroke lesions need to be fully segmented with accurate boundaries.
This work introduces the Multi-Stage Cross-Scale Attention (MSCSA) mechanism, applied to the U-Net family, to improve the mapping between brain structural features and lesions of varying sizes. Using the Anatomical Tracings of Lesions After Stroke (ATLAS) v2.0 dataset, MSCSA outperforms all baseline methods in both Dice and F1 scores on a subset focusing on small lesions, while maintaining competitive performance across the entire dataset. Notably, the ensemble strategy incorporating MSCSA achieves the highest scores for Dice and F1 on both the full dataset and the small lesion subset.
These results demonstrate the effectiveness of MSCSA in segmenting small lesions and highlight its robustness across different training schemes for large stroke lesions. 
Our code is available at: \url{https://github.com/nadluru/StrokeLesSeg}.
\end{abstract}

\section{Introduction}
\label{sec:intro}

Strokes are caused by an insufficient supply of blood to certain parts of the brain. The localization and quantification of the injured tissue (lesion), as well as the ``penumbra'' (the at-risk peripheral tissue) using Magnetic Resonance Imaging (MRI), is fundamental to providing a prognosis and developing individualized recovery and rehabilitation plans. The task of manually tracing the lesion on MRI is tedious, time-consuming, prone to errors, and always requires input from trained radiologists who rarely have the time to provide guidance on a case-by-case basis. The complexity of stroke lesions, which can vary in size and be dispersed across multiple regions of the brain, further complicates this task. Segmentation of these lesions is not only crucial for understanding stroke impact but is also vital for improving MRI workflows, such as structural, functional, and diffusion imaging pipelines, and for building accurate recovery models.

Automating stroke lesion segmentation using deep learning, especially with U-Net~\cite{unet} architectures, has gained considerable attention. The U-Net comprises an encoder and a decoder, constructed with convolutional blocks connected by skip connections across various resolutions. The encoder processes input images into latent features, which are subsequently decoded to generate the output segmentation mask. Concurrently, the attention mechanism~\cite{attention} has gained prominence in computer vision. Vision Transformers (ViT)~\cite{vit} emerged as a state-of-the-art model surpassing traditional convolutional networks and effectively modeling relationships between image patches over large distances.

Several pioneering works~\cite{hatamizadeh2021swin,hatamizadeh2022unetr} have explored integrating the encoder-decoder structure of U-Net with the long-distance modeling capabilities of Vision Transformers (ViT) in medical imaging. Building on these efforts, this study aims to advance stroke lesion segmentation by proposing a plug-in module that modifies the skip connections in U-Net, benefiting the entire U-Net family.
To achieve this, we introduce the Multi-Stage Cross-Scale Attention (MSCSA) module~\cite{mscsa} into the U-Net architecture. MSCSA facilitates multi-stage interaction by concatenating outputs from different convolution blocks in the encoder and creating multi-stage feature maps. This interaction enables MSCSA to establish connections between representative and abstract feature maps from different layers in the network to learn heterogeneous lesions present in different brain regions. Furthermore, MSCSA achieves cross-scale interactions by processing features at different scales and generating a cross-scale attention map, thus enabling MSCSA to effectively learn relationships between objects of different sizes, a crucial capability given that stroke lesions in a brain can be a collection of distinct components that vary in size.

\begin{figure*}[!h]
    \centering
    \includegraphics[scale = 0.314]{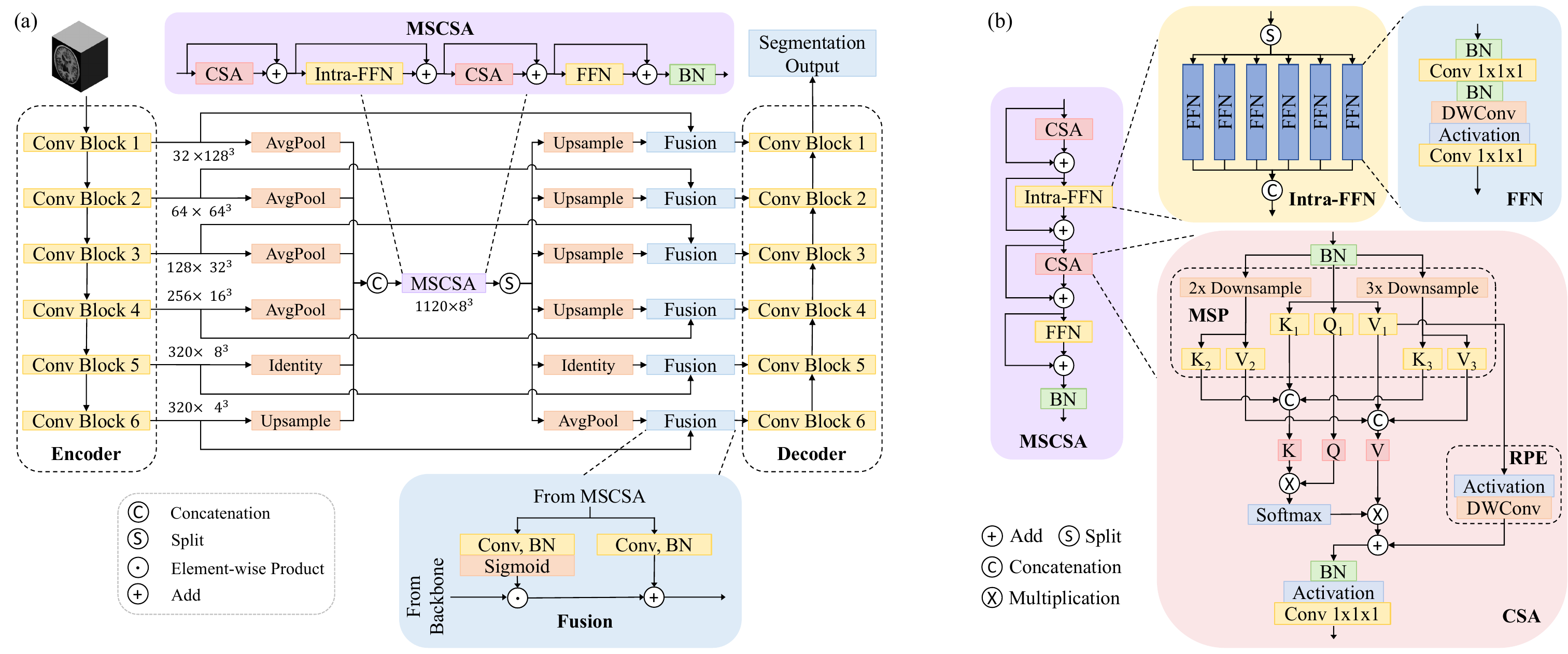}
    \caption{\textbf{(a) Multi-Stage Cross-Scale Attention (MSCSA) as an add-on module.} Feature maps from each Convolution (Conv) Block in the encoder are down(up)sampled to a uniform scale, concatenated along the channel dimension, and directed into the MSCSA block. The resulting output feature maps after a Batch Normalization (BN) layer undergo a reversal process, reverting to their original shapes. These refined maps are subsequently fused with feature maps sourced from the encoder and transmitted to the decoder. \textbf{(b) Cross-Scale Attention (CSA) and Intra-Feed-Forward Network (Intra-FFN).} In addition to the standard attention mechanism, CSA introduces the Multi-Scale key and value Projection (MSP) and Relational Positional Encoding (RPE). The MSP manipulates the key and value tensors across three scales, culminating in the generation of a cross-scale attention map. The RPE, which includes an activation layer and a Depth-Wise Convolution (DEConv) layer, enhances local feature aggregation through positional encoding. For Intra-FFN, feature maps are segmented into distinct parts corresponding to the number of channels in each encoder stage. Following individual processing through the standard Feed-Forward Network (FFN), these segmented maps are concatenated, reinstating their multi-stage format.}
    \label{fig:overview}
\end{figure*}

To assess the performance of U-Net enhanced with MSCSA, we conducted extensive experiments on the Anatomical Tracings of Lesions After Stroke (ATLAS) v2.0 dataset~\cite{atlas}. This includes employing various schemes for a comprehensive comparison with the baseline U-Net, encompassing the standard training pipeline, modifications to the loss function, adoption of a more complex architecture, implementation of a self-training scheme \cite{mapping}, and the exploration of an ensemble scheme. Our findings indicate that MSCSA exhibits superior performance for smaller lesions while remaining competitive with all the baseline schemes for large lesions.

\section{Methods}
\label{sec:method}

\noindent\textbf{Multi-Stage Cross-Scale Attention (MSCSA).}
Our framework builds upon the MSCSA~\cite{mscsa} and nnU-Net~\cite{mapping,nnunet}, allowing seamless integration with the broader U-Net family. As depicted in {\bf Fig.}~\ref{fig:overview}(a), the MSCSA serves as an add-on module, replacing direct connections between the U-Net's encoder and decoder. Feature maps of varying resolutions from the encoder are first adjusted to a target size of $H\times W\times D$, then concatenated along the channel dimension to create multi-stage feature maps. These are then processed through the MSCSA block, which includes a Cross-Scale Attention (CSA) layer followed by an Intra-Feed-Forward Network (Intra-FFN) layer and another CSA layer followed by a Feed-Forward Network (FFN) layer. Post-processing involves splitting and up(down)sampling to the original resolution. Output feature maps are converted into injection weights and biases and fused with feature maps from the encoder. The resulting feature maps, enriched with multi-stage and cross-scale information and spatial information from the encoder, are sent to the decoder to generate segmentation masks.

\noindent\textbf{Cross-Scale Attention (CSA).} 
Compared to the widely recognized attention mechanism, our proposed Cross-Scale Attention (CSA) integrates additional cross-scale interaction through the utilization of the Multi-Scale key and value Projection (MSP) and incorporates positional encoding via the Relational Positional Encoding (RPE), illustrated in {\bf Fig.}~\ref{fig:overview}(b).

Within the MSP, the input multi-stage feature maps with a resolution of $h \times w \times d$ generate three distinct branches with varying scales $h_{i} \times w_{i} \times d_{i}$ for $i=1,2,3$ as given by
\vspace{-0.75mm}
\begin{gather}\small
    \begin{aligned}
        h_{1} &= h,\! &h_{2} &= \left\lfloor \frac{h - 1}{2} + 1 \right\rfloor,\! &h_{3} &= \left\lfloor \frac{h - 1}{3} + 1 \right\rfloor, \\
        w_{1} &= w,\! &w_{2} &= \left\lfloor \frac{w - 1}{2} + 1 \right\rfloor,\! &w_{3} &= \left\lfloor \frac{w - 1}{3} + 1 \right\rfloor, \\
        d_{1} &= d,\! &d_{2} &= \left\lfloor \frac{d - 1}{2} + 1 \right\rfloor,\! &d_{3} &= \left\lfloor \frac{d - 1}{3} + 1 \right\rfloor. \\
    \end{aligned}
\vspace{-1.75mm}
\end{gather}
Similar to a standard attention layer, feature maps in the original scale are transformed into the query tensor $\Q$, key tensor $\K_{1}$, and value tensor $\V_{1}$ with a size of $h_{1}w_{1}d_{1} \times c_{j}$ for $j=q, k, v$, respectively, where $c_{j}$ is the dimensions of the features. Meanwhile, in the downsampled branches, additional key tensors $\K_{2}$, $\K_{3}$, and value tensors $\V_{2}$, $\V_{3}$ are generated. Here,  $\K_{i}$ has a size of $h_{i}w_{i}d_{i} \times c_{k}$ and $\V_{i}$ has a size of $h_{i}w_{i}d_{i} \times c_{v}$ for $i=2,3$. The key and value tensors across these three different scales are then concatenated, forming the multi-scale key tensor $\K$ and value tensor $\V$ via
\vspace{-0.75mm}
\begin{equation}\small
    \begin{aligned}
        \K &= \text{Concat}(\K_{1},\K_{2},\K_{3}) 
        ,\\
        \V &= \text{Concat}(\V_{1},\V_{2},\V_{3}) 
        .
    \end{aligned}
    \label{eq:kv_concat}
\vspace{-1.75mm}
\end{equation}
The concatenated $\K$ and $\V$ have dimensions of $(h_{1}w_{1}d_{1} + h_{2}w_{2}d_{2} + h_{3}w_{3}d_{3}) \times c_{k}$ and $(h_{1}w_{1}d_{1} + h_{2}w_{2}d_{2} + h_{3}w_{3}d_{3}) \times c_{v}$, respectively.
Subsequently, the query tensor $\Q$ in the original scale collaborates with the multi-scale key and value tensors $\K$, $\V$ in a cross-scale attention operation following
\vspace{-1mm}
\begin{equation}\small
    \text{Attn}(\Q, \K, \V) = \text{Softmax}\left(\frac{\Q\K^{T}}{\sqrt{c_{k}}}\right)\V \in \R^{h_{1}w_{1}d_{1} \times c_{v}}.
    \label{eq:attn}
\vspace{-1mm}
\end{equation}

While the attention mechanism models long-distance relationships, the short-distance relationship between nearby voxels is also crucial for 3D image data. Inspired by~\cite{hrvit}, we introduce a direct connection between the value tensor $\V_{1}$ in the original scale and the attention output, facilitated by an activation function and a Depth-Wise Convolution (DWConv) layer. The incorporation of DWConv amplifies local feature aggregation and serves as a relational positional encoding.

\noindent\textbf{Intra-Feed-Forward Network (Intra-FFN).}
As shown in {\bf Fig.}~\ref{fig:overview}(a), the multi-stage feature map directed to MSCSA comprises 1120 channels. Given such a large number of channels, employing linear layers (Conv \(1\times1\times1\)) in FFN, which further expands the channel count threefold, would result in significant computational complexity. To mitigate this, we implement Intra-FFN to alleviate the computational bottleneck. Within the Intra-FFN, the multi-stage feature maps are partitioned based on the number of stages in the encoder. The channel dimension of each segment matches that of the respective encoder stage. Each segment undergoes independent processing through a standard FFN. The resulting output feature maps are then concatenated and reintegrated as multi-stage feature maps. To balance the information gained from global interactions and the computational complexity, half of FFNs were replaced with Intra-FFNs.

\noindent\textbf{Dataset}
Our models were evaluated on the Anatomical Tracings of Lesions After Stroke (ATLAS) v2.0 dataset~\cite{atlas}, consisting of a training set of 655 T\(_1\)-weighted MRIs with corresponding lesion segmentation masks. All images were spatially normalized and registered to the MNI-152 template~\cite{mazziotta2001probabilistic}. Lesion volumes in this dataset range from 13 to over 200,000 voxels. To specifically assess performance on small lesions, we selected 138 MRIs with lesion volumes under 1,000 voxels to create a dedicated testing subset.

\noindent\textbf{Implementation details.}
Our models were built using PyTorch~\cite{pytorch} and nnU-Net~\cite{nnunet}. The training was conducted over 1000 epochs, employing a batch size of 2, Stochastic Gradient Descent (SGD) as the optimizer with an initial learning rate of 0.01 and a momentum value of 0.99. For model training, we adopted a 5-fold size-balanced cross-validation approach~\cite{mapping}, wherein the entire training dataset was partitioned into five folds, each ensuring a nearly uniform distribution of lesion sizes. Throughout the training process, image patches of dimensions $128 \times 128 \times 128$ were extracted from the original MR images as input to the model.

\noindent\textbf{Training schemes.}
To evaluate the performance of MSCSA against baseline models, we conducted experiments using four different approaches. First, we integrated MSCSA into a standard U-Net architecture, using a combined loss function of Dice score and Cross-Entropy (CE) loss. Next, we replaced CE loss with a Top10 CE loss (DTK10) to specifically enhance performance for smaller lesions. In the third approach, we swapped the standard U-Net with a Res U-Net to test MSCSA's effectiveness with a more advanced backbone. Additionally, we applied a self-training strategy following the methodology in~\cite{mapping}, where pseudo-masks for lesions in an additional 300 unlabeled MRIs from the test set were generated using the best model from the first three approaches. A generic U-Net was then trained using the original training dataset combined with these predicted pseudo-lesions. Finally, we employed an ensemble method by averaging the softmax outputs of all models from the four schemes.

\section{Results}
\label{sec:result}

\begin{table}[t]
    \centering
    \begin{tabular}{l|cc|cc}
        \toprule
        \multirow{2}{*}{Method} 
        & \multicolumn{2}{c|}{Entire} 
        & \multicolumn{2}{c}{Small Lesion} 
        \\ \cline{2-5} 
        & \makebox[0.084\linewidth][c]{Dice}
        & \makebox[0.084\linewidth][c]{F1}
        & \makebox[0.084\linewidth][c]{Dice}
        & \makebox[0.084\linewidth][c]{F1}
        \\
        \midrule
        Default
        & 0.635
        & 0.549
        & 0.417
        & 0.500
        \\
        \rowcolor{mygray}
        Default+MSCSA
        & 0.636
        & 0.551
        & 0.436
        & 0.527
        \\
        \midrule
        DTK10
        & 0.629
        & 0.547
        & 0.416
        & 0.491
        \\
        \rowcolor{mygray}
        DTK10+MSCSA
        & 0.629
        & 0.539
        & 0.417
        & 0.505
        \\
        \midrule
        Res U-Net
        & 0.638
        & 0.540
        & 0.433
        & 0.502
        \\
        \rowcolor{mygray}
        Res U-Net+MSCSA
        & 0.633
        & 0.546
        & 0.447
        & 0.528
        \\
        \midrule
        Self-Training
        & 0.647
        & 0.550
        & 0.434
        & 0.510
        \\
        \rowcolor{mygray}
        Self-Training+MSCSA
        & \textbf{0.648}
        & 0.553
        & 0.456
        & 0.534
        \\
        \midrule
        Ensemble
        & 0.646
        & 0.568
        & 0.439
        & 0.537
        \\
        \rowcolor{mygray}
        Ensemble+MSCSA
        & \textbf{0.648}
        & \textbf{0.573}
        & \textbf{0.458}
        & \textbf{0.574}
        \\
        \bottomrule
    \end{tabular}
    \caption{\textbf{Evaluation results of the 5-fold cross-validation.} MSCSA demonstrates competitive performance across all baseline methods on the full dataset and significantly outperforms the corresponding baselines on the small lesion subset in all metrics. Notably, the Ensemble+MSCSA approach achieves the highest performance on both the entire dataset and the small lesion subset.}
    \label{tab:dice}
\end{table}

\begin{figure*}[t]
    \vspace{-7mm}
    \hspace{-9mm}
    \includegraphics[scale=0.53]{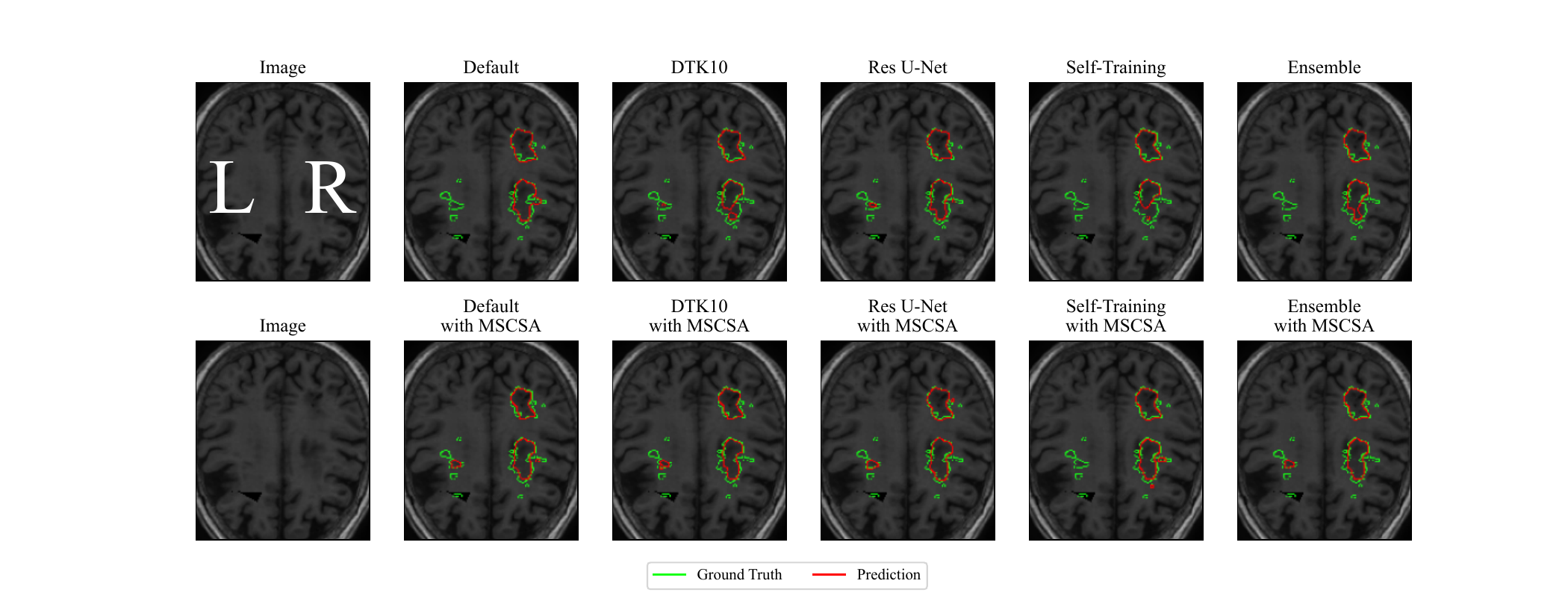}
    \vspace{-9.5mm}
    \caption{\textbf{Visualization of lesion segmentation results.} All the methods work well on larger lesion areas (see lesions in the right hemisphere of the brain), while MSCSA generalizes better for small lesions (see lesions in the left hemisphere).}
    \label{fig:visual}
\end{figure*}

\begin{figure*}[t]
    \centering
    \vspace{-2mm}
    \includegraphics[width=0.99\linewidth]{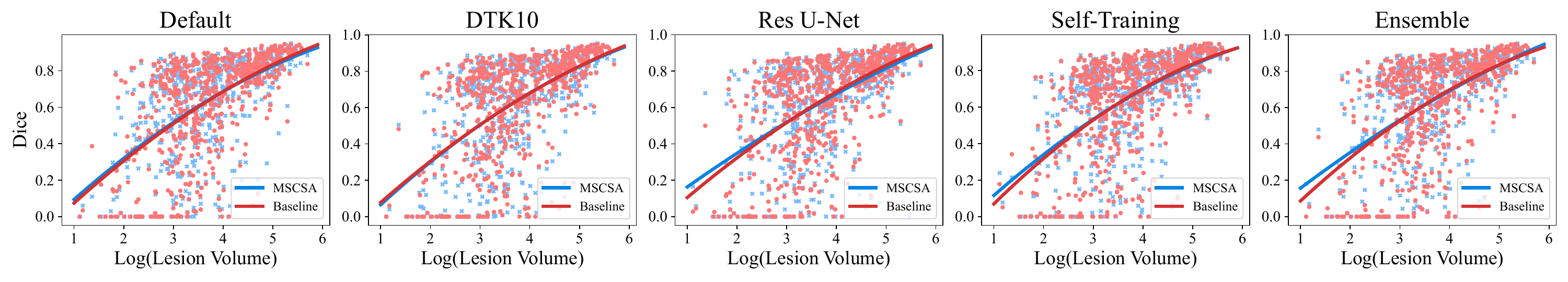}
    \vspace{-6mm}
    \caption{\textbf{Dice score vs. stroke lesion volume.} We can observe that MSCSA (blue) surpasses the baseline (red) for small lesions while simultaneously being competitive with the baseline on medium and large lesions.}
    \label{fig:dice_volume}
    \vspace{-3mm}
\end{figure*}

\noindent\textbf{Performance on the entire dataset}
We first present the evaluation of MSCSA using 5-fold cross-validation on the entire dataset, as outlined in Table~\ref{tab:dice}. In the Default, DTK10, Res U-Net, and Self-Training schemes, MSCSA demonstrates competitive and comparable performance with the baselines. On the other hand, in the ensemble schemes, MSCSA emerges as the superior method on both the Dice and F1 scores.

\noindent\textbf{Performance on the small lesion subset} 
The proposed Multi-Stage Cross-Scale Attention (MSCSA) integrates multi-stage and cross-scale interactions of image features, enabling a more effective mapping of brain structural features to lesion characteristics and an enhanced ability to detect even small lesions. Hence, we conducted another series of evaluations on 138 MRIs from the entire dataset that only have lesions with volumes under 1000 voxels. As shown in ~\ref{tab:dice}, MSCSA surpasses all corresponding baseline methods across all metrics. Notably, Ensemble+MSCSA achieves the highest Dice and F1 scores.

Moreover, as shown in Fig.~\ref{fig:visual}, in the right hemisphere of the brain, which contains a few medium-sized lesions, both the baseline and MSCSA perform well in identifying the lesion area. However, in the left hemisphere of the brain with smaller lesions, MSCSA demonstrates superior generalization ability and consistently identifies the small lesion areas across most schemes. Additionally, Fig.~\ref{fig:dice_volume} depicts the relationship between Dice scores and lesion volumes across different schemes. In all the schemes except for DTK10, the fitting curve indicates that MSCSA excels in detecting small lesions while remaining competitive for large lesions.  

The evaluation results in Table~\ref{tab:dice}, along with the visual representations in Fig.~\ref{fig:visual} and the curves relating Dice scores and lesion volumes in Fig.~\ref{fig:dice_volume}, collectively affirm the effectiveness of MSCSA in detecting small lesions without compromising its performance on large lesions.

\section{Discussion}
\label{sec:conclu}
This study integrated the Multi-Stage Cross-Scale Attention (MSCSA) module into the U-Net family for stroke lesion segmentation from T\(_1\) weighted MRI data. MSCSA demonstrated improved efficacy in detecting and segmenting small lesions while maintaining competitive performance with a wide variety of training schemes for large lesions. 
The proposed method can be further refined with advanced data augmentation strategies, such as Multi-Size Labeling (MSL) and Distance-Based Labeling (DBL)~\cite{shang2024segmenting}. These techniques involve segmenting training labels into connected components of varying sizes or defining them based on the distance between lesion and non-lesion voxels, enabling independent and more targeted training. By capturing size- and shape-specific features effectively, these strategies enhance MSCSA's generalization and applicability for MRI-based stroke lesion characterization.
We also plan to extend our framework to broader applications, such as Multiple Sclerosis (MS) lesion segmentation~\cite{rondinella2024icpr}, where lesions typically fall below 1000 $\text{mm}^{3}$, further demonstrating its versatility and clinical utility.

\section{Acknowledgements}
\label{sec:ack}
The following NIH grants are acknowledged: R01NS123378, R01NS111022, R01NS105646, R01NS117568, and \\P50HD105353.
\renewcommand*{\thefootnote}{\arabic{footnote}}
\section{Compliance with ethical standards}
This research study was conducted retrospectively using human subject data (ATLAS v2.0~\cite{atlas}) made available in open access under the Creative Commons Attribution 4.0 International License\footnote{\url{http://creativecommons.org/licenses/by/4.0/.}}. Ethical approval was not required as confirmed by the license attached with the open access data.

\bibliographystyle{IEEEbib}
\bibliography{refs}

\begin{thebibliography}{10}

\bibitem{unet}
Olaf Ronneberger, Philipp Fischer, and Thomas Brox,
\newblock ``U-net: Convolutional networks for biomedical image segmentation,''
\newblock in {\em Medical Image Computing and Computer-Assisted Intervention--MICCAI 2015: 18th International Conference, Munich, Germany, October 5-9, 2015, Proceedings, Part III 18}. Springer, 2015, pp. 234--241.

\bibitem{attention}
Ashish Vaswani, Noam Shazeer, Niki Parmar, Jakob Uszkoreit, Llion Jones, Aidan~N Gomez, {\L}ukasz Kaiser, and Illia Polosukhin,
\newblock ``Attention is all you need,''
\newblock {\em Advances in neural information processing systems}, vol. 30, 2017.

\bibitem{vit}
Alexey Dosovitskiy, Lucas Beyer, Alexander Kolesnikov, Dirk Weissenborn, Xiaohua Zhai, Thomas Unterthiner, Mostafa Dehghani, Matthias Minderer, Georg Heigold, Sylvain Gelly, et~al.,
\newblock ``An image is worth 16x16 words: Transformers for image recognition at scale,''
\newblock {\em arXiv preprint arXiv:2010.11929}, 2020.

\bibitem{hatamizadeh2021swin}
Ali Hatamizadeh, Vishwesh Nath, Yucheng Tang, Dong Yang, Holger~R Roth, and Daguang Xu,
\newblock ``Swin unetr: Swin transformers for semantic segmentation of brain tumors in mri images,''
\newblock in {\em International MICCAI brainlesion workshop}. Springer, 2021, pp. 272--284.

\bibitem{hatamizadeh2022unetr}
Ali Hatamizadeh, Yucheng Tang, Vishwesh Nath, Dong Yang, Andriy Myronenko, Bennett Landman, Holger~R Roth, and Daguang Xu,
\newblock ``Unetr: Transformers for 3d medical image segmentation,''
\newblock in {\em Proceedings of the IEEE/CVF winter conference on applications of computer vision}, 2022, pp. 574--584.

\bibitem{mscsa}
Liang Shang, Yanli Liu, Zhengyang Lou, Shuxue Quan, Nagesh Adluru, Bochen Guan, and William~A Sethares,
\newblock ``Vision backbone enhancement via multi-stage cross-scale attention,''
\newblock {\em arXiv preprint arXiv:2308.05872}, 2023.

\bibitem{atlas}
Sook-Lei Liew, Bethany~P Lo, Miranda~R Donnelly, Artemis Zavaliangos-Petropulu, Jessica~N Jeong, Giuseppe Barisano, Alexandre Hutton, Julia~P Simon, Julia~M Juliano, Anisha Suri, et~al.,
\newblock ``A large, curated, open-source stroke neuroimaging dataset to improve lesion segmentation algorithms,''
\newblock {\em Scientific data}, vol. 9, no. 1, pp. 320, 2022.

\bibitem{mapping}
Jiayu Huo, Liyun Chen, Yang Liu, Maxence Boels, Alejandro Granados, Sebastien Ourselin, and Rachel Sparks,
\newblock ``Mapping: Model average with post-processing for stroke lesion segmentation,''
\newblock {\em arXiv preprint arXiv:2211.15486}, 2022.

\bibitem{nnunet}
Fabian Isensee, Paul~F Jaeger, Simon~AA Kohl, Jens Petersen, and Klaus~H Maier-Hein,
\newblock ``nnu-net: a self-configuring method for deep learning-based biomedical image segmentation,''
\newblock {\em Nature methods}, vol. 18, no. 2, pp. 203--211, 2021.

\bibitem{hrvit}
Jiaqi Gu, Hyoukjun Kwon, Dilin Wang, Wei Ye, Meng Li, Yu-Hsin Chen, Liangzhen Lai, Vikas Chandra, and David~Z Pan,
\newblock ``Multi-scale high-resolution vision transformer for semantic segmentation,''
\newblock in {\em Proceedings of the IEEE/CVF Conference on Computer Vision and Pattern Recognition}, 2022, pp. 12094--12103.

\bibitem{mazziotta2001probabilistic}
John Mazziotta, Arthur Toga, Alan Evans, Peter Fox, Jack Lancaster, Karl Zilles, Roger Woods, Tomas Paus, Gregory Simpson, Bruce Pike, et~al.,
\newblock ``A probabilistic atlas and reference system for the human brain: International consortium for brain mapping (icbm),''
\newblock {\em Philosophical Transactions of the Royal Society of London. Series B: Biological Sciences}, vol. 356, no. 1412, pp. 1293--1322, 2001.

\bibitem{pytorch}
Adam Paszke, Sam Gross, Francisco Massa, Adam Lerer, James Bradbury, Gregory Chanan, Trevor Killeen, Zeming Lin, Natalia Gimelshein, Luca Antiga, et~al.,
\newblock ``Pytorch: An imperative style, high-performance deep learning library,''
\newblock {\em Advances in neural information processing systems}, vol. 32, 2019.

\bibitem{shang2024segmenting}
Liang Shang, Zhengyang Lou, Andrew~L Alexander, Vivek Prabhakaran, William~A Sethares, Veena~A Nair, and Nagesh Adluru,
\newblock ``Segmenting small stroke lesions with novel labeling strategies,''
\newblock in {\em International Workshop on Machine Learning in Clinical Neuroimaging}. Springer, 2025, pp. 113--122.

\bibitem{rondinella2024icpr}
Alessia Rondinella, Francesco Guarnera, Elena Crispino, Giulia Russo, Clara Di~Lorenzo, Davide Maimone, Francesco Pappalardo, and Sebastiano Battiato,
\newblock ``Icpr 2024 competition on multiple sclerosis lesion segmentation—methods and results,''
\newblock in {\em International Conference on Pattern Recognition}. Springer, 2024, pp. 1--16.

\end{thebibliography}

\end{document}